# Non-Gaussian Dephasing of an Electronic Mach-Zehnder Interferometer Coupled to "Which Path" Detector

I. Neder

*Braun Center for Submicron Research, Department of Condensed Matter Physics,*

*Weizmann Institute of Science, Rehovot 76100, Israel*

**A theoretical non-pertubative treatment is developed to explain the dephasing of electrons in the electronic Mach-Zehnder interferometer via interaction with a near-by partitioned electronic channel, which acts as a "which path" detector. The resulting formula reproduces the recant experimental behavior of the MZI interference visibility. By fitting the model to the experimental results, it is shown that the visibility is strongly influenced by merely ~3 detecting electrons, hence it reflects the Non-Gaussian properties behavior of the detector shot-noise.**

## A. Introduction

"Controlled dephasing" experiments are used to study the transition from quantum to classical behavior (the vanishing of the interference effects) in coherent mesoscopic systems. In past experiments an electron interferometer was coupled to another quantum device (a 'which path detector'), acting as the environment [1-2]. Since in those experiments the coupling between the detector and the interferometer was small, dephasing was not perfect, and resulted from many, weakly detecting electrons. Under these conditions, the phase of the interfering electron can be described semi-classically by a Gaussian like random process [3-6].

A more recent controlled dephasing experiment was preformed on an electronic Mach-Zehnder Interferometer (MZI) [7,8], using a near by partitioned edge-channel [9,10]. It showed substantially different results compared to the controlled-dephasing experiments in the past. The main purpose of this paper is to provide a theoretical model that explains these mew experimental results. However the system analyzed here is quiet fundamental – decoherence and orbital entanglement [11] of strongly interacting two 1-dimensional electronic channels. Hence, the model and the calculations below stand on their own right as a scheme of solving this problem non-perturbatively.

A simplified scheme of the experiment is presented in Fig. 1, and was explained thoroughly in Ref. 9,10. Both the MZI and the detector were realized utilizing chiral 1-dimentional edge-channels in the integer Quantum Hall effect regime. The MZI phase was controlled by modulation gate via the Aharonov Bohm effect [12]. The additional edge channel was partitioned by a quantum point contact (**QPC0**) and the reflected part flew in close proximity to the upper path of the MZI, serving as "which path" phase-sensitive detector [2]. As the bias in the detector channel increased the Coulomb interaction caused orbital entanglement between the interfering electron and the detecting electrons passing by which scrambled the phase of the interfering electron, causing classical behavior.

The contrast of the AB oscillations (the visibility of the MZI) was measured as a function of the DC bias on the detector channel - $V_{det}$, and the partitioning of the detector channel $T_{QPC0}$. Three new and peculiar effects were observed:

a) The visibility behavior as a function of $T_{QPC0}$ changed from a smooth parabolic behavior at low $V_{det}$ (small dephasing) to a rather sharp V-shape behavior at large $V_{det}$, with almost total dephasing at $T_{QPC0}$=0.5 (see Fig. 3 in Ref. 9)
b) Non-monotonic behavior of the visibility with increasing $V_{det}$: For some values of **QPC0** gate voltages, The AB oscillations dropped to zero at $V_{det}$~14µV, then reappeared again as $V_{det}$ increased, and vanished finally at $V_{det}$~30 µV (see Fig 4 in Ref. 9). For some other gate voltages it decreases monotonically (see Fig. 2 in Ref. 10).
c) The system has a unique noise property, which allows recovering the phase information, after it has completely vanished in conductance measurements, by cross-correlating the current fluctuations of the MZI and detector [10].

We suspected that these effects may be a signature of strong dephasing by only few (1-3) detecting electrons [10]. The dephasing in our system is caused by the quantum shot-noise of the detecting electrons [13] which scrambles the phase of the interfering electron [14]. This noise is not Gaussian, but has a *binomial* nature. Because the detector is very sensitive and the coupling between the two systems is strong, the higher-moments of this noise becomes important, so treating it perturbatively up the second moment is not good enough. In other words, the dephasing depends on the noise's full counting statistics, which have been given much attention in recant years [15-20].

In Ref 9. we showed that a simple model of detecting with one discrete electron state (that have a binomial shot-noise statistics) provides a qualitative explanation to the experimental results (a)-(c), but with clear quantitative shortcomings. The natural explanation for these shortcomings is that detection in the experiment is due to more then one electron. Then two questions arise: (1) How many electrons dephase the MZI as $V_{det}$ increases, and (2) how is the dephasing effect distributed between the detecting electrons. Moreover, the notion "detection of one electron by few electrons" itself demands clarification, as both the MZI and the detector are open long chiral 1-dimentional channels (edge channels) which contain many electrons in *extended* quantum states.

This paper presents a solvable model that answers the questions above, and provides quantitative predictions for the behavior of the visibility with detector bias and partitioning. It is organized as follows. In section B the Hamiltonian is presented, which describes the interfering electron, the detecting electrons, and the coupling between them. The Schrödinger equation is solved, giving expression for the full wave-function. In section C the measured visibility is identified as an expectation value of a unitary "phase operator". This expectation value is calculated in section D, using some approximations based on physical reasoning, which allows us to throw away all the redundant degrees of freedom and be left with a finite dimensional phase operator that can be diagonalized numerically. Some general properties of the resulting formula are discussed in section E. These are used in section F in the comparison with the experimental result. In section G the Gaussian limit of the anzatz is explored. The conclusions are given in Section H.

**B. A model for the MZI+detector**

We consider here a quantum model for the interaction between an interfering electron in the MZI and electrons in a partitioned, biased, 1-dimentional channel – a detector. The MZI source has only a small bias on it, so the current is low and allows for only one interfering electron at a given time. Therefore the single-particle picture (first quantization) is assumed to be adequate (We will assume zero temperature throughout the paper). This assumption is not so trivial and remains to be clarified elsewhere, because we neglect Pauli blocking; the affect of other electrons preventing the interfering electron of using energy states lower then the Fermi energy [21,22]. The detector has many electrons in it, so it must be treated in second quantization. In our experiment we create entanglement using the one-particle state in the MZI and the many-body state in the detector.

Note in Fig 1 that the x-axis is defined along the MZI's upper arm, with x=-L defined at **QPC1** and x=0 at **QPC2**. Electrons in the detector are partitioned by **QPC0**, with probability R$_{QPC0}$ to be reflected and to flow in parallel and in close proximity to the electron in the MZI. We shell use the same x-axis for the detector's reflected part.

We consider the MZI electron single-particle wave function $|\psi\rangle$, That is coupled to the many body wave function of the electrons in the detector $|\Psi_{det}\rangle$, through the Hamiltonian

$$H = \varepsilon_M(\hat{p}) + \sum_k \varepsilon_{det}(\hbar k) c_k^+ c_k + \int dx' \hat{\rho}_{det}(x',t) u(x'-\hat{x}) \quad (1)$$

where $\hat{x}, \hat{p} \equiv -i\hbar \frac{\partial}{\partial x}$ act on the MZI 1-particle space, $c_k^+$ and $c_k$ are creation and annihilation operators of the ingoing detector k state from source S$_{det}$ (The other source is grounded and a zero temperature is assumed), $\varepsilon_M$ and $\varepsilon_{det}$ are the 1d dispersions in the MZI and detector channel respectively, u(x) is the coulomb inter-channel interaction between the electron in the MZI and the one in the detector and $\hat{\rho}_{det}(x) = \Psi_{det}^{r+}(x)\Psi_{det}^r(x)$ is the electron density operator in the reflected part of the detector.

Using the interaction picture for the detector Hilbert space, we first find the detector incoming state. Due to partitioning by **QPC0** and the detector bias V$_{det}$, the incoming energy states in the range $E_F < \varepsilon < E_F + eV_{det}$ are partitioned, and the detector state is

$$|\Psi_{det}\rangle = \prod_{\varepsilon(k)=E_f}^{E_f+eV} c_k^+ |\Psi_0\rangle = \prod_{\varepsilon(k)=E_f}^{E_f+eV} \left( t_{QPC0} c_k^{t+} + i r_{QPC0} c_k^{r+} \right) |\Psi_0\rangle, \quad (2)$$

where $r_{QPC0}$, $t_{QPC0}$ are **QPC0** reflection and transmission amplitudes, and $c_k^{r+}$ and $c_k^{t+}$ are the **outgoing** creation operators for the reflected and transmitted part, satisfying $c_k^+ = t_{QPC0} c_k^{t+} + i r_{QPC0} c_k^{r+}$. It is this partitioning that makes $\hat{\rho}_{det}(x,t)$ fluctuate, which leads to the scrambling of the phase and the loss of the interference pattern in the MZI, which we want to quantify.

We focus on the partial wave function $|\psi(x,t)\rangle \equiv \langle x|\Psi_{total}(t)\rangle$, which is the projection of the full wave function (MZI+detector) of the system onto the MZI single particle state $|x\rangle$ (Generally, the full wave-function can always be written as $|\Psi_{total}(t)\rangle = \int dx |x\rangle \otimes |\psi(x,t)\rangle$). It is a state in the detector Hilbert space, whose norm $\langle\psi(x,t)|\psi(x,t)\rangle$ gives the probability amplitude of the MZI electron to be in place x. Its Schrödinger equation is given by

$$i\hbar \frac{\partial}{\partial t}|\psi(x,t)\rangle = \left(\varepsilon(\hat{p}) + \hat{V}(x,t)\right)|\psi(x,t)\rangle, \qquad (3)$$

where $\hat{V}(x,t) \equiv \int dx' \hat{\rho}_{det}(x',t) u(x'-x)$. Note that the same equation would describe the 1-particle MZI wave-function, if it was subjected to some **classical** density fluctuations in the detector. However in our case the interfering electron senses the detector state and influences it at the same time, so entanglement occurs.

The system dynamics is influenced only by energy states some ~20μV around the Fermi energy, hence we are allowed to linearize the dispersions of Eq. 1-3 [21]. By introducing $|\psi(x,t)\rangle = e^{ik_F x - i w_F t} \cdot |\varphi(x,t)\rangle$, and Tailor expanding $\varepsilon(\hat{p})$ to first order, the equation for $|\varphi(x,t)\rangle$ now reads

$$i\hbar \frac{\partial}{\partial t}|\varphi(x,t)\rangle = \left(-i\hbar v_M \frac{\partial}{\partial x} + \hat{V}(x,t)\right)|\varphi(x,t)\rangle \qquad (4)$$

where $v_M$ is the Fermi velocity in the MZI. The same linearization can be done with the detector dispersion, resulting in a simple solution for the density operator:

$\hat{\rho}_{det}(x,t) = \tilde{\rho}_{det}(x - v_{det}t)$, where $v_{det}$ is the detecting electrons Fermi velocity, which can differ in principle from $v_M$.

Eq. 4 has an *exact solution* corresponding to every incoming k-wave $|\varphi_{k0}(x,t)\rangle = e^{i(kx - v_M t)} \cdot |\Psi_{det}\rangle$, which is

$$|\varphi_k(x,t)\rangle = T \exp\left[ikx - iv_M kt - \frac{i}{\hbar v_M}\int^x dx' \cdot \hat{V}\left(x', t - \frac{x-x'}{v_M}\right)\right]|\Psi_{det}\rangle, \quad (5)$$

where T denote time ordering. This means that at a given time and place (x,t), the phase of the wave function is contributed by an integral of all values of $\hat{V}(x',t')$ at those previous times and places (x',t') that lies on the "influence line" $x - x' = v_M(t - t')$. $|\varphi_k(x,t)\rangle$ contains all the information about the entanglement between the MZI electron and the detector electrons.

## C. Identifying the measured visibility

Before biasing the detector, we measured the interference in the MZI by splitting the 1-particle wave at x=-L (**QPC1**) to upper part and lower part $\psi_u$ and $\psi_l$ respectively, and recombining the two parts at x=0 (**QPC2**), with an AB phase shift that can be controlled during the experiment. The phase-dependant part of $|\psi_{D2}|^2$ at drain **D2** is proportional to $\text{Re}(\psi_l^*(0,t) \cdot \psi_u(0,t))$, and we have to average this over time. After averaging, this term oscillates as $v \cdot \cos(\phi_{AB})$ with $\phi_{AB}$ the AB phase, and $v$ the observed visibility, which turns out to be the magnitude of the average,

$$v = \lim_{T \to \infty} \left|\left(\frac{1}{T}\int_0^T \psi_l(x,t)^* \cdot \psi_u(x,t)dt\right)\right|.$$

We will assume for brevity that the initial visibility is 100%, and the reduction is only due to detection. In reality the visibility starts from some initial value due to external fluctuations [21,22] (35-65% in the actual experiment in Ref. 7-10) and the dephasing is measured in percentage relative to this value.

Note that following the definition of $|\psi(x,t)\rangle$, also $|\psi_u\rangle$ and $|\psi_l\rangle$ belongs to the detector Hilbert space, and when the detector is working, they carries different detector wave functions, so the visibility is now the average over the scalar product [3,4]:

$$v = \lim_{T\to\infty}\left|\left(\frac{1}{T}\int_0^T \langle\psi_l(0,t)|\psi_u(0,t)\rangle dt\right)\right| \qquad (6)$$

Now $|\psi_l\rangle \propto e^{-iw_F t}|\Psi_{det}\rangle$ oscillates with a single energy component. Using this in Eq. 6, we get that the visibility is just the Fourier component of $|\langle\Psi_{det}|\psi_u(0,t)\rangle|$ at the initial energy – *which is the square root of the probability for an electron to exit the MZI with the same energy it entered*; not to receive any real energy from the detector and not to change the detector initial state. We are going to use this fact later on.

using Eq. 5, in Eq. 6, with $|\psi_u\rangle = e^{-i(kx-v_M t)}|\varphi_k(x,t)\rangle$, we get

$$v = \left|\left\langle\Psi_{det}\left|T\exp\left[\frac{i}{\hbar v_M}\int_{-L}^0 dx' \cdot \vec{V}\left(x', t - \frac{x-x'}{v_F}\right)\right]\right|\Psi_{det}\right\rangle\right|. \qquad (7)$$

We can drop time average since the problem is stationary, so we take t=0 in all the calculations below.

The last step is to represent Eq. 7 as an expectation value of a unitary operator

$$v = \left|\langle \Psi_{det} | e^{i\hat{\Phi}} | \Psi_{det} \rangle\right|, \tag{8}$$

by dropping of the time ordering operator. We are allowed to do so because the density operator is Bosonic in one dimension – $[\hat{\rho}(x), \hat{\rho}(x')]$ is a c-number. The time order exponent in Eq. 7 is by definition a product of many small unitary evolutions sorted by time. Hence, using repeatedly the known relation $e^A e^B = e^{A+B} e^{\frac{1}{2}[A,B]}$ which holds for any two Bosonic operator A and B, we can collect the operators at different times together in the same exponent. The remaining c-number exponent (the part $e^{\frac{1}{2}[A,B]}$) has unit magnitude, so it does not lead to reduction in the visibility and we can disregard it.

The *phase operator* $\hat{\Phi}$ in Eq. 8 is therefore a weighted integral over the density operator at time x=t=0:

$$\hat{\Phi} \equiv \frac{1}{\hbar v_M} \int_{-L}^{0} dx' \cdot \hat{V}\left(x', t - \frac{x-x'}{v_F}\right)_{x=t=0} = \int_{-\infty}^{\infty} w(x) \tilde{\rho}_{det}(x) dx. \tag{9}$$

The weight $w(x)$ is positive definite and is found by inserting the definitions of $\hat{V}(x,t)$ and $\tilde{\rho}(x)$ above into the first integral of Eq. 9 and rearranging the integrals;

$$w(x) = \left| \frac{1}{\hbar(v_M - v_{det})} \int_{\frac{v_{det}-v_M}{v_M}L}^{0} u(x-y) dy \right|, \tag{10}$$

which is a convolution of $u(x)$ with the "window of influence" defined by the MZI path length L and the ratio of the velocities.

**D. Diagonalization of the phase operator**

We believe that the fluctuations of the phase are highly non-Gaussian, so we shall not expand Eq. 8 in moments (for we cannot take only the second one). Instead, we follow the traditional QM route: We shell find $\hat{\Phi}$'s eigenvalues and eigenstates. Then we represent the detector state $|\Psi_{det}\rangle$ in the eigenstates basis, and take the average. The only problem is that as $\hat{\Phi}$ is defined, it is infinite dimensional. Our task is to reduce $\hat{\Phi}$ into a finite matrix, by identifying those degrees of freedom that contribute to the dephasing. This can be achieved in the following two step process.

First, note that only real **down**-scattering process of the detector energy contributes to the dephasing. As stated in Section C, dephasing is induced by energy transfer between the detector and the MZI. The electron in the MZI has no energy to give (its energy is near the Fermi surface), hence Pauli principle rules out the contribution of up-scattering processes above $E_F+eV_{det}$. Those up-scattering events exist as virtual processes and their effect is probably to renormalize parameters such as the velocities $v_M, v_{det}$ and to screen the Coulomb interaction $u(x)$.

Following this line of thought, we make here an approximation, by assuming that we can work with the renormalized values and *redefine* the density in Eq. 9, *restricting its excitations to the voltage window*:

$$\tilde{\rho}_{det}(x) = \frac{1}{L_{det}} \sum_{\varepsilon(k)=0}^{eV_D} \sum_{\varepsilon(k+q)=0}^{eV_D} c_k^{r+} c_{k+q}^r e^{iqx} \tag{11}$$

(Up-scattering within [0, eV] must remain, to make $\tilde{\rho}(x)$ Hermitian). $L_{det}$ is the length of the whole detector channel – and will be cancelled out in the end result.

There are still many states between $E_F$ and $E_F+eV_D$ (for a long detector channel). Our second step is to move from momentum (k) basis to the real-space basis, by construction of wave-packets from the states within the above energy range [23]:

$$\psi_n(x-v_Dt) = \frac{1}{\sqrt{hv_{det}eV_{det}}} \int_{E_F}^{E_F+eV_{det}} e^{i\frac{E}{\hbar v_{det}}[x-v_{det}(t-n\tau_{det})]} dE, \quad -\infty < n < \infty, \quad (12)$$

where $\tau_{det} = h/eV_{det}$ is the time interval between two successive wave packets ($\psi_n(x-v_{det}\tau_{det}) = \psi_{n-1}(x)$). This set of "(sin x)/x" wave packets is orthonormal basis for the detector 1-P states. The n'th wave function is localized around $x_n = v_{det} \cdot (t - n\tau_{det})$. In second quantization, the annihilation operator of this state is

$$\hat{\psi}_n(t) = \sqrt{\frac{hv_{det}}{eV_{det}L_{det}}} \cdot \sum_{\varepsilon(k)=E_F}^{E_F+eV_{det}} e^{ik\cdot v_{det}(t-n\tau_{det})} \hat{c}_k \, , \text{ satisfying the usual anti-commutation relation}$$

$$\{\hat{\psi}_n(t), \hat{\psi}_m^+(t)\} = \delta_{mn}.$$

Diagonalization in this basis is now much easier. Note that the phase operator, defined by Eq. 8-11, does not change the number of particles, but only mixes the states inside the energy interval $[E_F, E_F+eV]$. Hence, we can write the one-particle matrix elements of $\hat{\Phi}$ using the wave-packet basis:

$$\Phi_{nm} \equiv \langle \psi_m | \hat{\Phi} | \psi_n \rangle = \langle 0 | \hat{\psi}_m \hat{\Phi} \hat{\psi}_n^+ | 0 \rangle = \frac{hv_{det}}{eV_{det}L_{det}^2} \sum_{\varepsilon(k)=E_F}^{E_F+eV_{det}} \sum_{\varepsilon(k')=E_f}^{E_F+eV_{det}} e^{iv_{det}\tau(kn-k'm)} w_{k'-k} \quad (13)$$

where $w_q$ is the Fourier transform of $w(x)$, $w_q = \int_{-\infty}^{\infty} w(x)e^{iqx}dx$. Note that although n and m runs from $-\infty$ to $\infty$, $\{\Phi_{nm}\}$ is usually "localized" in the sense that for large |n| or |m| the wave-packets lies "outside" of the influence region w(x) and therefore those matrix elements are negligible. Hence one can ignore "far-away" elements to make the matrix finite, and diagonalize it numerically.

Diagonalizing the one-particle matrix $\{\Phi_{nm}\}$ is all we need for calculating the visibility. when we find the eigenvalues $\phi_i$ and the normalized eigenvectors $\bar{v}_i$, we can write the phase operator as $\hat{\Phi} = \sum_{i=-\infty}^{\infty} \phi_i \hat{v}_i^+ \hat{v}_i$, where $\hat{v}_i, \hat{v}_i^+$ are the electron annihilation and creation operators at the i's (one particle) eigenstate of $\hat{\Phi}$: $\hat{v}_i = \sum_n (\bar{v}_i)_n \hat{\psi}_n$, which satisfies $\{\hat{v}_i, \hat{v}_j^+\} = \delta_{ij}$. The operator $\hat{v}_i^+ \hat{v}_i$ is just the occupation of the i's state. It commutes with any other state occupation $\hat{v}_j^+ \hat{v}_j$, and with respect to $|\Psi_{\text{det}}\rangle$ all those occupation operators have probability $R_{QPC0}$ to have the value 1 and probability $T_{QPC0}$ to have the value 0. Taking this into account, Eq. 8 reduces to

$$v = \left|\langle e^{i\hat{\Phi}} \rangle\right| = \left|\langle \Psi_{\text{det}} | \prod_{i=-\infty}^{\infty} e^{\phi_i \hat{v}_i^+ \hat{v}_i} | \Psi_{\text{det}} \rangle\right| = \left|\prod_i \left(R_{QPC0} e^{i\phi_i(V_{\text{det}})} + T_{QPC0}\right)\right| \qquad (14)$$

This formula resembles expressions that were obtained before [20] regarding full counting statistics of electrons.

Equations 13 and 14 are the main results of the paper. They give close expression of the visibility of the AB oscillations in the MZI as a function of detector bias and partitioning. Given $V_{det}$, there are specific electron states $|v_i\rangle$ in the detector whose occupation influence the MZI phase by different amounts $\phi_i$. These occupations fluctuate according to the partitioning at **QPC0.** The visibility is the product of all those influences.

Before investigating Eq. 13,14, there are two simplifications that one can always perform on Eq. 13 which provides better understanding and also makes the numerical calculations much easier. First, it is generally worthwhile to convert the sums in Eq. 13 into integrals, using the conversion $\sum_k = \frac{L_{det}}{2\pi} \int dk$, And also to switch to the dimensionless variables using the wave packets width $\tau_{det} v_{det}$: $\tilde{x} = x/\tau_{det} v_{det}$, $\tilde{k} = \tau_{det} v_{det}(k - k_F)/2\pi$. We shell also use a normalized weighting function $\tilde{w}(\tilde{x}) = \tau_{det} v_{det} w(x) / \int w(x) dx$. After these adjustments, Eq. 13 reads,

$$\Phi_{mn} = \gamma V_{det} \cdot \int_0^1 \int_0^1 e^{i 2\pi (\tilde{k}n - \tilde{k}'m)} \tilde{w}_{\tilde{k}'-\tilde{k}} d\tilde{k} d\tilde{k}', \qquad (15)$$

Where $\tilde{w}_{\tilde{q}}$ is the Fourier transform of $\tilde{w}(\tilde{x})$ and $\gamma \equiv \frac{e}{h v_{det}} \int w(x) dx$. This form, while very useful for numerical purposes, may be misleading, as one should notice that the integrals still depend on the detector bias, through the definition of $\tilde{w}_{\tilde{q}}$ - increasing the bias makes the scale of this function "shrink" relative to the finite borders of the integrals.

The second simplification is achieved by performing one of the two integrals in Eq. 15. This can be done by switching to a new variables of integration - $\tilde{k}, \tilde{k}' \to \tilde{k}, \tilde{q} = \tilde{k}' - \tilde{k}$, and make the integral over $\tilde{k}$. The borders of integration for $\tilde{q}$ are -1 and 1, so the borders of $\tilde{k}$ become $\max(0, -\tilde{q})$ and $1 - \max(0, \tilde{q})$. Calculating the integral over $\tilde{k}$ we get (after many straightforward adjustments that I skip here) the final result:

$$\Phi_{mn} = \gamma V_{\text{det}} \begin{cases} \int_{-1}^{1} d\tilde{q} \left[ e^{-2\pi i \tilde{q} m} (1 - |\tilde{q}|) w(\tilde{q}) \right] & m = n \\ \dfrac{1}{\pi(m-n)} \int_{-1}^{1} d\tilde{q} \left[ e^{-\pi i \tilde{q}(m+n)} \sin(\pi(n-m)|\tilde{q}|) w(\tilde{q}) \right] & m \neq n \end{cases} \qquad (16)$$

Which looks more complicated then Eq. 15, but is much easier to evaluate numerically.

**E. General properties of the visibility**

We start exploring the behavior of the visibility, with the case of low detector bias $V_{det}$, or alternatively, a very localized $w(x)$. There exist a finite range of small $V_{det}$, in which $w_k$ stays roughly constant (equals to $w_0$) in the relevant range $0 < k < \dfrac{eV_D}{\hbar v_D}$. In that case, Eq. 15 reduces to:

$$\Phi_{mn} = \gamma V_{\text{det}} \delta_{m0} \delta_{n0}. \qquad (17)$$

This means that our choice of basis (The wave-packets basis) diagonalizes the phase operator; if the "zero" wave packet is occupied with an electron it has the eigenvalue $\gamma V_{\text{det}}$, and if the wave packet is unoccupied it has the eigenvalue 0. The occupation probability is again $R_{QPC0}$. Tracing over the occupation of other wave packets will contribute nothing. We get the final result according to Eq. 14:

$$v = \left| \langle e^{i\hat{\Phi}} \rangle \right| = \left| R_{QPC0} e^{i\gamma V_{\text{det}}} + T_{QPC0} \right|, \qquad (18)$$

This is the result of the "one detecting state" approximation [7,8]. It may at first look highly counterintuitive; detection by a *single* electron state, whose phase is bias dependent. Naively one would assume [1-6] that each passing electron in the detector induces a constant phase shift to the interfering electron, and changing the detector current (via changing the bias $V_{det}$) controls how many electrons are passing in the detector. In fact, it is the *exactly* the opposite! The number of detecting states remains one because of Pauli principle – at low $V_{det}$ only one state can enter the small "influence window" w(x). The linear phase dependence on $V_{det}$ can be understood by energy consideration: phase scrambling costs energy, and the detecting electron has only a limited energy $eV_{det}$ to give away.

At higher bias more detecting states influence and the one detecting state approximation, Eq. 18, looses its validity. Technically, this is because w(x) has a finite width – lets call it $\Delta$. For large enough $V_{det}$, $\widetilde{\Delta} \equiv \frac{\Delta}{\tau_{det} v_{det}} > 1$ ($\widetilde{\Delta}$ is the width $\widetilde{w}$), and the matrix $\Phi_{mn}$ now has non-diagonal elements with respect to the wave-packet basis. There are now few significant eigen-values $\phi_i(V_{det})$ (their number increasing with $V_{det}$). For each $V_{det}$ one should construct $\Phi_{mn}$ and diagonalize it. Then for each set of eigenvalues $\{\phi_i(V_{det})\}$ the visibility is calculated according the Eq. 14.

We close this section by stating two important properties of the eigenvalues of the phase matrix:
1. For any detector bias $V_{det,}$ the eigenvalues obey a sum rule; $\sum_i \phi_i = \gamma W_{det}$. This can be easily obtained from Eq. 16 by calculating $tr\Phi$, (the trace equals to the sum of all eigenvalues). It can also be obtained by calculating for the case of a full beam ($R_{QPC0} = 1$) in Eq. 14 and comparing it with the experimental result of a constant phase shift slope with $V_{det}$ (Fig. 2 in Ref. 9). This means that the parameter $\gamma$ was actually measured in that experiment – we identify it with this slope of the MZI phase as a function of V$_{det}$.

2. There exist an important dimensionless, *bias independent* parameter in the problem, $N_\pi = \pi \frac{\widetilde{\Delta}(V_{det})}{\gamma V_{det}}$. Roughly it measures how many wave packets are needed to enters the influence region w(x) to shift the MZI phase by $\pi$. The larger $N_\pi$, the weaker is the interaction between the channels, and then dephasing is due many weakly interacting electrons. On the other hand when $N_\pi$ is small, the dephasing is due few strongly interacting electrons.

**F. comparison with the experimental results**

We compare first the one detecting state approximation; Eq 18. Since $\gamma$ was measured, Eq. 18 does not have free parameters, and can be compared directly with the experimental data. As shown in Fig. 3 in Ref. 9, it fits very well to the results at low $V_{det}$ and qualitatively it produces the effects a-c, in Section A. In particular, it is easy to show that it predicts the change in the behavior of the visibility dependence on T$_{QPC0}$ from a smooth shape to a V-shape, as well as the non-monotonous behavior with increasing $V_{det}$. However, according the Eq. 18 these effects happens near the value $\gamma V_{det} = \pi$, while the experimental results shows that the V-shape and the zero of the visibility happens at *larger detector bias (by ~40%)*! Hence quantitatively at this detector bias Eq. 18 fails to reproduce the experimental results.

The reason of this discrepancy must be onset of other detecting electrons. According to Eq. 18, the visibility should have a zero minima when $\phi_0 = \gamma V_{det} = \pi$. However if other eigenvalues become slightly non-zero, then $\phi_0$ becomes smaller then $\gamma W_{det}$ (because of the sum rule). The visibility then has its zero minima at *larger* $V_{det}$, which satisfies $\phi_0 = \pi$ (from Eq. 14). At even larger bias, near $\gamma_0 = 2\pi$ the visibility will have again maxima, however it will be smaller then 100% due to the dephasing of the other detecting states. These two effects has both been seen in the experiment (Fig 4. in Ref. 9). By taking a Lorntzian guess $\widetilde{w}_{\widetilde{q}} = \frac{1}{1+\widetilde{\Delta}^2 \widetilde{q}^2}$; $\widetilde{\Delta}(V_{det}) = \gamma W_{det} N_\pi / \pi$, and fitting the theory to the experimental results we find $N_\pi$=1.6. More practically, at $V_{det}$=9.5 ($\gamma W_{det} = \pi$), the first eigenvalue is $\gamma_0 \sim 0.8\pi$, which means strong inter-channel interaction, and entanglement involving almost a single pair of electrons.

**G. The Gaussian limit**

Even though in our experiment $N_\pi$ turned out at the order of unity, it is instructive to explore theoretically what happens when the interaction is very weak, and $N_\pi$ is very large – that is the Gaussian limit. In this limit any finite reduction of the visibility is due to many *weakly interacting* electrons, so all the phases $\{\phi_i\}$ are much smaller than π. The number of detecting electrons is defined roughly as the number of wave packets that enters the "influence region", $N \equiv \widetilde{\Delta} = N_\pi \frac{\gamma W_{det}}{\pi}$. The phases $\{\phi_i\}$ can then be obtained analytically from Eq. 15, up to zero order of $1/N$. We can change the limits of the integration of $\widetilde{q}$ to $[-\infty, \infty]$, for the correction is at the order of $\frac{1}{\widetilde{\Delta}} = \frac{1}{N}$ and can be neglected. In addition, for m=n, we can neglect the $|\widetilde{q}|$ term, and for $m \neq n$ the whole expression will be at the order of $1/N$, because of the sin function.

Summing it all up we have:

$$\Phi_{nm} = \gamma V_{det}\delta_{nm}\int_{-\infty}^{\infty} e^{i2\pi \tilde{q}n}\tilde{w}_{\tilde{q}}d\tilde{q} + O\left(\frac{1}{N}\right) \cong \gamma V_{det}\tilde{w}(n)\delta_{nm} = w(nv_{det}\tau_{det})\delta_{nm} \quad (19)$$

We get a result in the space domain, which is a very intuitive: the phase matrix in this case is again almost diagonal in the wave packet basis. The wave packets are small relative the width of w(x) and every wave-packet "peaks up" the phase according to the local value of w(x) at it's center - $\phi_i = w(v_D \tau_D i)$.

Turning now to compute the product in Eq. 14, and expecting a Gaussian form, we can define the visibility and phase as an exponent of some complex function $\lambda(V_D)$

$$\langle e^{i\Phi}\rangle = e^{\lambda(V_D)} = \prod_i \left(R_{QPC0}e^{i\phi_i(V_D)} + T_{QPC0}\right) \quad (20)$$

Taking the log of both sides, and expending each term in the sum up to second order in $\phi_i(V_D)$ we get

$$\lambda(V_D) = \sum_i \log\left(R_{QPC0}e^{i\gamma_i} + T_{QPC0}\right) \approx iR_{QPC0}\gamma V_D - \frac{1}{2}R_{QPC0}T_{QPC0}\sum_i \phi_i^2 \quad (21)$$

This is the expected result – the phase evolution reacts to the mean field charging in the detector channel - $R_{QPC0}\gamma V_{det}$, and the visibility is reduced by an exponent of the second moment of the shot-noise - $R_{QPC0}T_{QPC0}$ times some factor - $\frac{1}{2}\sum_i \phi_i^2$. This factor can be calculated approximately using Eq. 19, and turning the sum into an integral on n,

$$\frac{1}{2}\sum_i \phi_i^2 = \frac{1}{2}\int_{-\infty}^{\infty} w^2(nv_{det}\tau_{det})dn = \frac{1}{2v_{det}\tau_{det}}\int_{-\infty}^{\infty} w^2(x)dx \quad (22)$$

In order that this factor will make sense as "phase diffusion" [3,4], we want to arrange it to be $\frac{1}{2}\bar{\varphi}_1^2 N$, where $\bar{\varphi}_1 = \frac{\beta V_{det}}{N}$ is the average phase shift induced by occupying a single wave packet. Note that we still have some freedom in defining N, by choosing the proper definition of $\Delta$. Hence, for large N $\Delta$ is found (by comparing the RHS of Eq. 22 to $\frac{1}{2}\bar{\varphi}_1^2 N$) to be:

$$\Delta = \frac{\left(\int_{-\infty}^{\infty} w(x)dx\right)^2}{\int_{-\infty}^{\infty} w^2(x)dx} \quad (23)$$

which indeed gives a faithful value for the width of the w(x).

**H. conclusions**

This work presented a quantum model to controlled dephasing experiment of interfering electron coupled to a biased and partitioned 1d channel. The interaction between the two channels causes the electrons in them to be in an orbital entangled state. The interaction in the experiment was strong, in the sense that only 1-2 detecting electron were sufficient to scramble the phase of the interference by π; this is the meaning of the parameter $N_\pi$. The strong interaction forces us to treat the problem non-pertubatively, in term of a "phase matrix" that must be diagonalized in order to find the proper detecting electron states and their influence. Exact formula of the visibility was derived which is valid, under some general and physically reasonable assumptions, for all dephasing range.

In the two opposite limits of this formula, namely the limit of dephasing with one strongly coupled electron ($N_\pi \ll 1$), and the limit of dephasing using many weakly coupled electrons ($N_\pi \gg 1$), the phase matrix is diagonalized using the Martin-Landauer wave-packets basis, giving simple results. However as in our case $N_\pi \sim 1.6$, neither of these two limits can't explain the experimental results. This means that there is no way to avoid the full resulting formula (Eq. 13,14) of this model. Lorenzian guess for $\widetilde{w}_{\widetilde{q}}$ fits fairly well to the experimental data and gives the above estimation for $N_\pi$. Assuming more realistic functional form for $w_{\widetilde{q}}$ may yield a better fit, and would probably change this estimation slightly, but can not change the main conclusion: the dephasing in our device results from entanglement between 1-3 electrons, and as such is sensitive to the full counting statistics of the quantum shot-noise.

**Acknowledgements**

I wish to thank F. Marquardt, E. Ginossar, Y. Levinson and M. Heiblum for hours of long and useful discussions, in which this work was checked and refined.

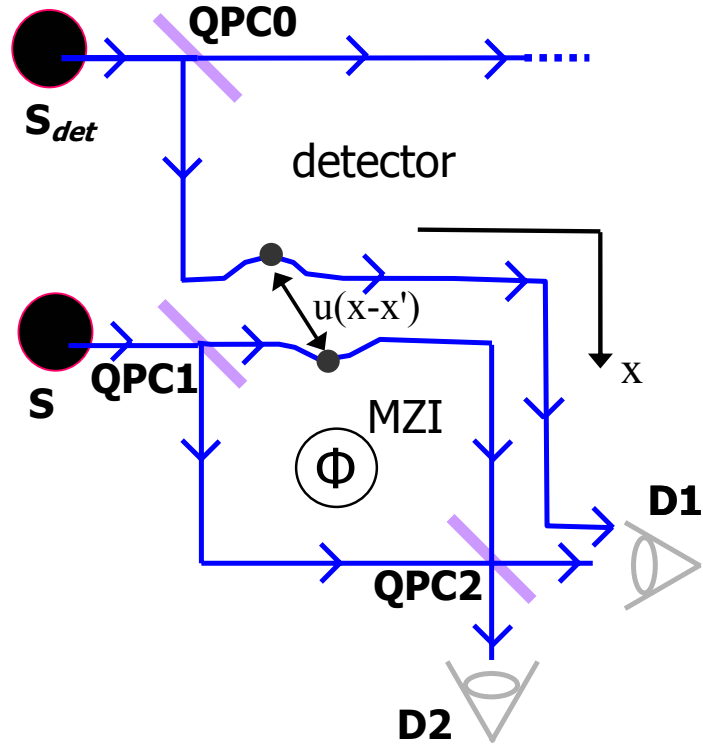

**Figure 1**: The electronic MZI and the phase-sensitive which path detector. The interference in the MZI causes the transmission probability from source S to drain D2 to be phase dependant and to oscillate when changing the AB flux Φ. an electron in the detector is partitioned by QPC0 and its reflected part passes near the MZI upper path and senses the coulomb repulsion from the interfering electron. The interaction causes a phase shift in both wave functions. Because of this phase shift the detector states contain information about the interfering electron being in the upper arm, hence detection. On the other hand, the detector's shot-noise causes scrambling of the phase of the interfering electron, leading to the loss of the interference pattern.